\documentclass[twocolumn]{aastex62}

\newcommand{\msol}{{\rm M}_{\rm \odot}} 

\received{May 11, 2018}
\revised{September 12, 2018}
\accepted{September 25, 2018}
\submitjournal{ApJ}

\shorttitle{Formamide in a fragmenting disc}
\shortauthors{Qu\'enard et al.}

\begin{document}

\title{THE FATE OF FORMAMIDE IN A FRAGMENTING PROTOPLANETARY DISC}

\correspondingauthor{John D. Ilee}
\email{j.d.ilee@leeds.ac.uk}

\author[0000-0002-2969-3985]{David Qu\'enard}
\affil{School of Physics and Astronomy, Queen Mary University of London, Mile End Road, London E1 4NS, UK}

\author[0000-0003-1008-1142]{John D. Ilee}
\affil{School of Physics \& Astronomy, University of Leeds, Leeds LS2 9JT, UK}
\affil{Institute of Astronomy, University of Cambridge, Madingley Road, Cambridge CB3 0HA, UK}

\author[0000-0003-4493-8714]{Izaskun Jim\'{e}nez-Serra}
\affiliation{Centro de Astrobiolog\'{i}a (CSIC/INTA), Ctra. de Torrej\'{o}n a Ajalvir km 4, 28850, Torrej\'{o}n de Ardoz, Spain}
\affiliation{School of Physics and Astronomy, Queen Mary University of London, Mile End Road, London E1 4NS, UK}

\author[0000-0003-1175-4388]{Duncan H. Forgan}
\affiliation{SUPA, School of Physics and Astronomy, University of St Andrews, North Haugh, St Andrews KY16 9SS, UK}
\affiliation{St Andrews Centre for Exoplanet Science, University of St Andrews, St Andrews KY16 9SS, UK}

\author[0000-0002-8138-0425]{Cassandra Hall}
\affiliation{Department of Physics and Astronomy, University of Leicester, Leicester LE1 7RH, UK} 

\author[0000-0002-8138-0425]{Ken Rice}
\affiliation{SUPA, Institute for Astronomy, University of Edinburgh, Blackford Hill, Edinburgh EH9 3HJ, UK}
\affiliation{Centre for Exoplanet Science, University of Edinburgh, Edinburgh EH9 3HJ, UK}

\begin{abstract}
Recent high-sensitivity observations carried out with ALMA have revealed the presence of complex organic molecules (COMs) such as methyl cyanide (CH$_3$CN) and methanol (CH$_3$OH) in relatively evolved protoplanetary discs.  The behaviour and abundance of COMs in earlier phases of disc evolution remains unclear.  
Here we combine a smoothed particle hydrodynamics simulation of a fragmenting, gravitationally unstable disc with a gas-grain chemical code.  We use this to investigate the evolution of formamide (NH$_{2}$CHO), a pre-biotic species, in both the disc and in the fragments that form within it.
Our results show that formamide remains frozen onto grains in the majority of the disc where the temperatures are $<$100 K, with a predicted solid-phase abundance that matches those observed in comets.  Formamide is present in the gas-phase in three fragments as a result of the high temperatures ($\geq$200\,K), but remains in the solid-phase in one colder ($\leq$150\,K) fragment.  The timescale over which this occurs is comparable to the dust sedimentation timescales, suggesting that any rocky core which is formed would inherit their formamide content directly from the protosolar nebula.
\end{abstract}


\section{Introduction} \label{intro}

The origin of the prebiotic content at the surface of the primitive Earth remains unclear. It is postulated that a large quantity of prebiotic material might have been brought to Earth by comets and meteorites during its early formation \citep{caselli2012}. This hypothesis is supported by the detection of numerous pre-biotically relevant molecular species in comets such as Hale-Bopp \citep{bockelee-morvan2000}, Lemmon and Lovejoy \citep{biver2014}, and more recently 67P/Churyumov-Gerasimenko \citep{goesmann2015, altwegg2017}. 

\smallskip

Among these species is formamide (NH$_2$CHO), an important precursor of pre-genetic and pre-metabolic compounds such as nucleic acids, nucleobases, sugars and amino acids \citep[see][]{saladino2012}. The reservoir of complex organic molecules (COMs) --- such as formamide --- in comets is expected to derive from the previous stages of star formation. Formamide has indeed been detected in the hot envelopes around low-mass protostars \citep[e.g.][]{kahane2013}. The unrivalled sensitivity of the Atacama Large Millimeter Array (ALMA) has opened up the possibility to also study COM chemistry in protoplanetary discs. However, only two COMs have been detected so far in the gas-phase in these objects: methyl cyanide (CH$_3$CN, \citealp{oberg2015}) and methanol (CH$_3$OH, \citealp{walsh2016}). A large proportion of these COMs are thought to remain frozen onto dust grains, making their detection in the gas-phase challenging \citep[see, e.g.][]{walsh2014}.  

\smallskip

There have been several theoretical studies into the evolution and inheritance of chemistry during the youngest stages of protoplanetary discs.  Some have considered the chemical link between a surrounding envelope and an axisymmetric disc \citep[e.g.][]{visser2011, drozdovskaya2016, furuya2017}, finding that chemical composition depends on the route taken by material to the forming disc.  Other works have chosen to concentrate on the chemistry of isolated discs, modelling the full three-dimensional hydrodynamic evolution within the disc itself \cite[e.g.][]{ilee2011, evans2015, yoneda2016}, finding that dynamically-induced changes in temperature and density of the disc material can significantly influence the chemical composition.  To date, only one study has followed the time dependent chemical evolution of a disc that forms protoplanetary fragments \citep{ilee2017}, but this work did not consider the formation and evolution of complex molecules.  Therefore, the formation and evolution of COMs from the protostellar phase to the birth of a planetary system remains unclear.

\smallskip

In this paper, we investigate the fate of one such COM, formamide, during one of the earliest phases of disc evolution. We consider a massive, self-gravitating protostellar disc that undergoes fragmentation which we model using Smoothed Particle Hydrodynamics (SPH) coupled with a gas-grain chemical evolution code. We examine the abundance of formamide in the different environments of the disc (e.g. cold outer regions, regions undergoing shocks) and particularly in the four protoplanetary fragments formed during the simulation.  We show that our models broadly reproduce observations of COMs in comets, and that formamide may survive around the fragments long enough to be incorporated into protoplanets rocky cores, implying the possibility of direct inheritance of formamide from the protosolar epoch.

\section{Methodology} \label{methods}

\subsection{Hydrodynamic modelling}

We use the same radiation-hydrodynamic disc simulation as in \citet{ilee2017}, which is also `Run 2' from \citet{hall2017}.  Full details of the simulation set up are given in these works, and so we only briefly reiterate key aspects here.  We simulate a 0.25~$\msol$ disc, with a 1~$\msol$ central protostar.  The gas in the disc is represented by $4\times10^{6}$ SPH particles distributed between 10 and 100~au, with an initial surface density profile that scales as $\Sigma (r) \propto r^{-1}$ and an initial sound speed profile that scales as $c_{\rm s} \propto r^{-0.5}$.  We include the modified radiation transfer scheme of \citet{forgan2009} in which the gas is able to cool radiatively according to its local optical depth (calculated from the gravitational potential), and can exchange energy with neighbouring fluid elements via flux-limited diffusion.   

\smallskip

Four fragments are formed during this simulation (labelled in ascending radial distance from the central star as John, Paul, George and Ringo, see \citealp{ilee2017}). John and Paul are tidally disrupted at $t=3081$\,yr and $t=3514$\,yr, respectively, while George and Ringo survive until the end of the simulation.  John contains the most mass (10.3\,$M_{\rm Jup}$) but undergoes tidal disruption, George is the most massive surviving fragment (8.2\,$M_{\rm Jup}$), Ringo reaches 5.4\,$M_{\rm Jup}$ at the end of the simulation, and Paul is the least massive fragment, reaching 3.7\,$M_{\rm Jup}$ before being disrupted.

\subsection{Chemical modelling}

The chemical calculations are performed using \texttt{UCLCHEM} \citep{viti2004, holdship2017}, a gas-grain chemical evolution code. The gas-phase reactions are taken from the UMIST database \citep{mcelroy2013} with additional reactions from \citet{quenard2018}. The grain surface reaction network is the one presented in \citet{quenard2018}. Several thermal and non-thermal processes are considered on the grain surface, such as: diffusion via thermal hopping and quantum tunnelling \citep{hasegawa1992}, thermal desorption \citep{hasegawa1992}, cosmic rays desorption, direct-UV and secondary-UV desorptions \citep{mcelroy2013, holdship2017} and chemically reactive desorption \citep{minissale2016}.

\smallskip

The network contains 364 species (243 in the gas-phase and 121 on the grain surface) and 3446 reactions. The starting elemental abundances are the same as in \citet{quenard2018}. Binding energies of molecules are taken from \citet{wakelam2017} (see Table \ref{tableDes}).  We note that we do not consider three-body reactions, expected to be important in high-density environments, within our chemical model.  However, \citet{ilee2011} included three-body reactions in their study of the chemical composition of a massive disc, finding that they have a small effect on the molecular abundances (see their Section 3.3).

\begin{figure*}[!ht]
	\centering
	\includegraphics[width=\textwidth]{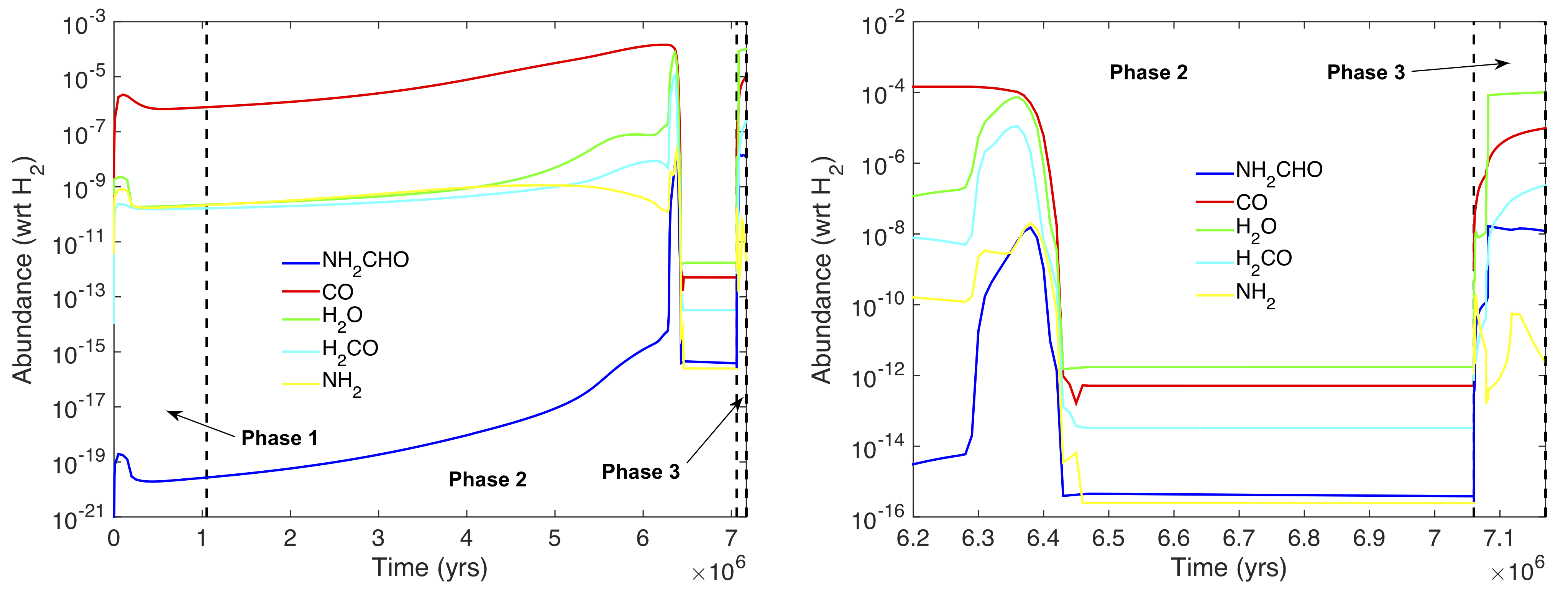}
	\caption{Gas-phase abundance of NH$_2$CHO (formamide), CO, H$_2$O, NH$_2$ and H$_2$CO across the pre-disc phases in our chemical modelling. \textit{Left panel:} Abundances across Phase 1--3 for the full $\sim$7.2\,Myr considered.  \textit{Right panel:} A close up of the final Myr showing Phases 2 \& 3. \vspace{0.5 cm}}
	\label{abund_prephases}
\end{figure*}

\begin{figure*}[!ht]
	\centering
	\includegraphics[width=0.49\hsize]{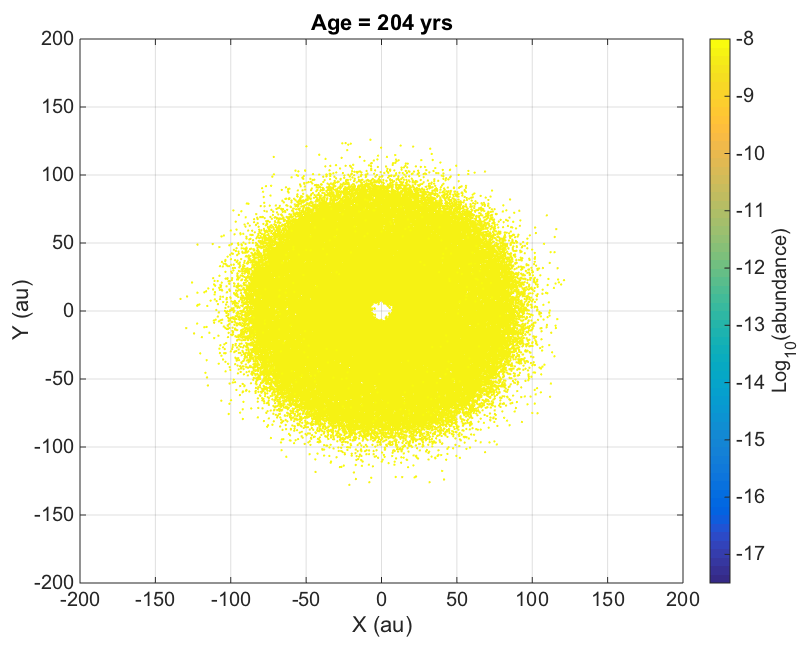}
	\includegraphics[width=0.49\hsize]{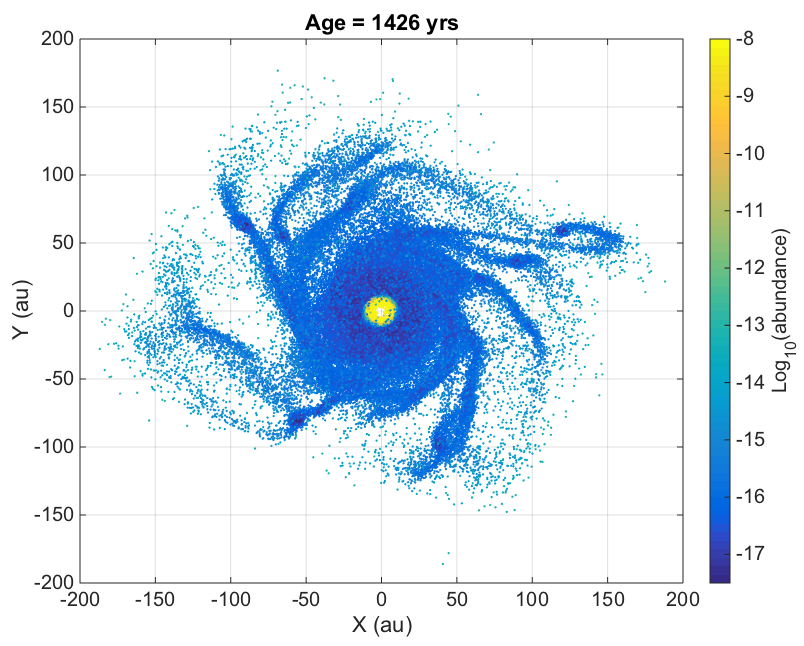}\\
	\includegraphics[width=0.49\hsize]{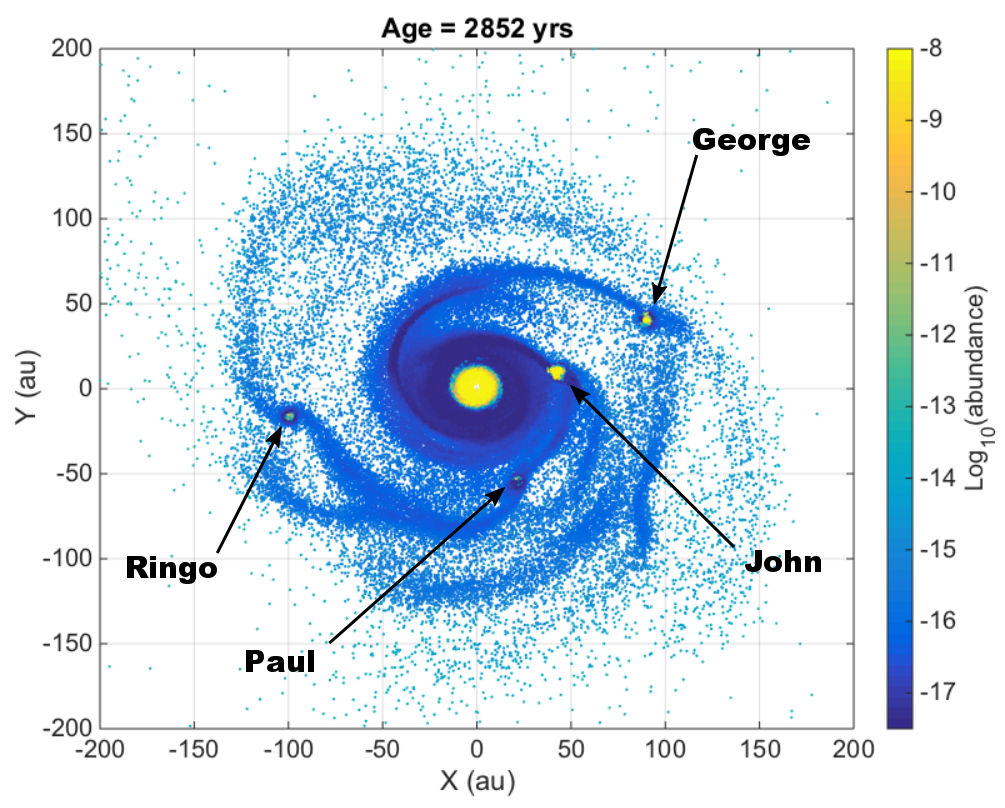}
	\includegraphics[width=0.49\hsize]{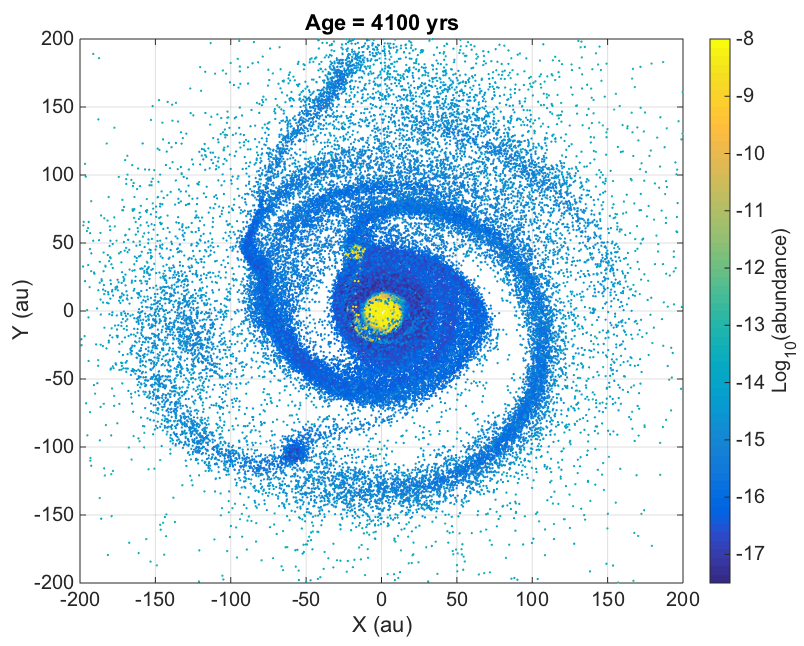}
	\caption{Gas-phase abundance of formamide across the protoplanetary disc for four different snapshots covering 4100\,yr of disc evolution. Formamide is kept on the grain surface in the majority of the disc but, with time, it goes back to the gas-phase in the central part of the disc and around some fragments. \vspace{0.5 cm}}
	\label{abund_formamide}
\end{figure*}

\smallskip

The chemical modelling is performed in four steps, where the first three initialise the chemical conditions prior to the full protoplanetary disc phase (i.e. the final abundances of phases 1, 2 and 3 are used as initial abundances for phases 2, 3 and 4, respectively).

\smallskip

The first step (Phase 1) corresponds to the \textit{diffuse cloud phase}, where we follow the chemistry of a low-density cloud (n$_{\rm H}=10^2$\,cm$^{-3}$ and A$_{\rm V}=2$\,mag) over 10$^6$\,yr at a kinetic temperature T$\rm _{kin}$=10\,K. Only gas-phase chemistry plays a role at this stage since the A$_{\rm V}$ is low, and all grain surface species are either destroyed or desorbed via non-thermal processes (see left panel of Fig. \ref{abund_prephases}).

\smallskip

The second step (Phase 2) is the \textit{pre-stellar phase}, where the cloud contracts for $\sim$5.4$\times10^6$\,yr until it reaches a final density of n$_{\rm H}=1\times10^{11}$\,cm$^{-3}$ following a free-fall collapse parametrisation \citep{rawlings1992, holdship2017}. The gas temperature remains at 10\,K during this phase. The cloud remains at low A$_{\rm V}\lesssim4$\,mag with n$_{\rm H}\lesssim2\times10^{3}$\,cm$^{-3}$ for the first $\sim$4.7$\times10^6$\,yr. During this period, the chemistry does not vary significantly and the abundances of formamide as well as those of species such as CO, H$_2$O, H$_2$CO and NH$_2$ remain relatively the same (see left panel of Fig. \ref{abund_prephases}). At approximately $6.3\times10^{6}$\,yr, almost all molecules experience a steep increase in gas-phase abundance.  As discussed in \citet{quenard2018}, the high density allows molecular species to be efficiently produced on grain surfaces since the high visual extinction prevents UV photons from destroying or desorbing molecules. These newly formed species are then injected into the gas phase via non-thermal desorption processes such as chemical reactive desorption. The abundances of H$_2$O, H$_2$CO, NH$_2$ and NH$_2$CHO drastically decrease after $6.4\times10^{6}$\,yr, as a result of the severe freezing out expected at the high density regime in pre-stellar cores.

\smallskip

The third step (Phase 3) is the \textit{warming-up (or protostellar/hot corino) phase}. While the density is kept constant and the temperature increases from 10 to 300\,K following a rate defined by \citet{awad2010}. Due to the increasing grain surface temperature, this phase enables the formation of COMs via radical-radical addition, and once the dust temperature exceeds the desorption energies shown in Table 1, molecular species are thermally desorbed into the gas-phase.  We also emphasise that higher temperatures also help to overcome the activation barrier of the endothermic reactions present in our gas-phase chemical network, which triggers a richer gas-phase chemistry.

\smallskip

The fourth step (Phase 4) is the \textit{protoplanetary disc phase}, when we run \texttt{UCLCHEM} for a subset of 100,000 particles taken from the SPH simulation. We proceed in this way to make it possible to undertake the calculation of the chemical evolution of the disc in a reasonable time-frame using the large network of UCLCHEM.  The subset of particles were selected randomly, but their selection was confined to within a smoothing-length of the mid-plane in order to preserve the radial surface density structure of the disc. The subset is large enough to ensure that all physical regimes of the disc are probed. For each particle within this subset, the starting abundance is taken to be the final abundance of the warming-up (proto-stellar) phase.  Since the subset is drawn from close to the mid-plane, we set the visual extinction (A$_{\rm V}$) in this phase to $>$100 mag, such that any photo-processing can be considered negligible.  For all phases, we assume an external radiation field of G$_0=1$\,Habing and a standard cosmic ray ionisation rate of $\zeta=1.3\times10^{-17}$\,s$^{-1}$.

\begin{table}[!ht]
	\centering
	\caption{Binding energies of formamide and related species adopted in our chemical model, taken from \citet{wakelam2017}. \label{tableDes}}
	\begin{tabular}{lcc}
		\hline\hline
		Species	&	Binding energy		\\
				&	E$_{\rm D}$\,(K)	\\
		\hline
		CO				&	1300		\\	
		CO$_2$			&	2600		\\
		CH$_4$			&	960		\\	
		HCN				&	3700		\\
		H$_2$O			&	5600		\\
		CH$_3$OH		&	5000		\\
		NH$_3$			&	5500		\\	
		NH$_2$			&	3200		\\	
		H$_2$CO			&	4500		\\
		NH$_2$CHO		&	6300		\\
		\hline
	\end{tabular}
\end{table}

\smallskip

We have benchmarked the chemical abundances obtained in our work with those calculated by \citet{ilee2017}, who used a smaller chemical network (125 species $-$ 89 in the gas, 36 on dust grains $-$ and 1334 reactions; see their Section 2.2). We find a good agreement between the two networks, especially for molecules which are key in the production of formamide (e.g. CO, CO$_2$, H$_2$CO, H$_2$O, NH$_2$ and NH$_3$), with abundances differing by less than a dex.  Such a result is reassuring given the large discrepancy in size between the two networks.   

\section{Results}

\subsection{Formamide (NH$_{2}$CHO) in the disc}\label{results}

Fig. \ref{abund_formamide} shows the abundance of formamide in four different time snapshots in the protoplanetary disc. The first snapshot at 204\,yr shows the initial high gas-phase abundance of formamide across the disc, inherited from the protostellar (hot corino) phase. Rapidly, around 850\,yr, all species are frozen-out onto dust grains because of the cold temperatures of the disc ($T$$\lesssim$20\,K) combined with high densities (n(H$_2$)$\gtrsim$10$^{10}$\,cm$^{-3}$). Later, formamide is released back into the gas-phase via thermal desorption in regions where $T$$\gtrsim$100\,K. The magnitude of these effects depends strongly on the position of the particle within the disc (e.g. whether they reside in shocked regions, or within fragments). The total amount of formamide (gas-phase plus solid-phase) does not vary significantly with time, since the bulk of formamide has already been produced during the protostellar phase. Therefore, the formamide abundance in the gas- or solid-phase is mainly driven by the temperature, leading either to freeze-out or to desorption.

\smallskip

In the innermost parts of the disc ($r<15$\,au), Fig. \ref{abund_formamide} shows that formamide is in the gas-phase. This is due to the high gas kinetic temperatures ($T\,=\,100$--400\,K) which thermally desorb formamide from grains. Note that the presence of UV and/or X-ray radiation fields from the central protostar are not taken into account here. Therefore, the abundance of formamide at $r<15$\,au might be lower due to the photo-destruction of this molecule. At larger radii, the high visual extinction (A$_{\rm V}$$\geq$470\,mag) associated with the large densities in the disc (up to 10$^{13}$\,cm$^{-3}$ in the midplane) will shield formamide from any radiation fields (both stellar and external).

\smallskip

The gas-phase formamide abundance in the outermost parts of the disc remains typically lower than 10$^{-12}$, even reaching $\sim$10$^{-18}$ in some regions. As the temperature is also low (close to $10$\,K), the bulk of formamide remains frozen onto dust grains (with a constant abundance of 10$^{-8}$) in the arms and the outer parts of the disc. In Section \ref{sec:comets}, we compare these solid-phase abundances of formamide in the outer disc predicted by our model with those measured observationally in Solar System comets.

\smallskip

\begin{figure*}[!ht]
\centering
\includegraphics[height=0.3\textwidth, trim={0.5cm 0cm 4cm 1.05cm},  clip]{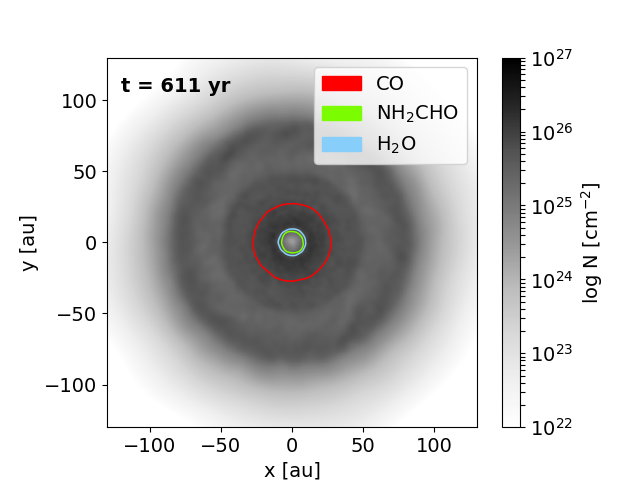}
\includegraphics[height=0.3\textwidth, trim={2.60cm 0cm 4cm 1.05cm},  clip]{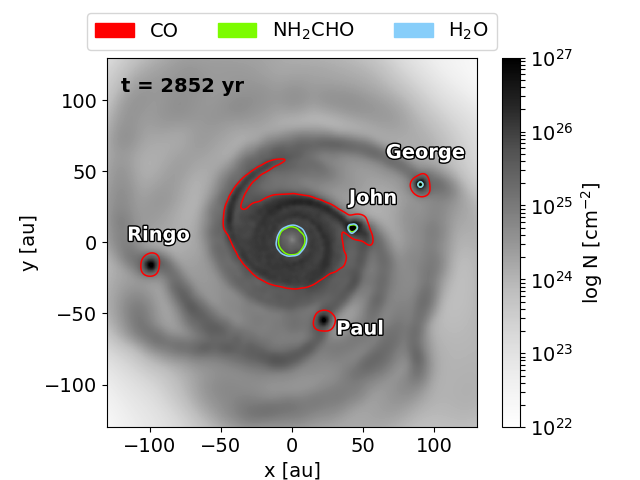}
\includegraphics[height=0.3\textwidth, trim={2.60cm 0cm 4cm 1.05cm},  clip]{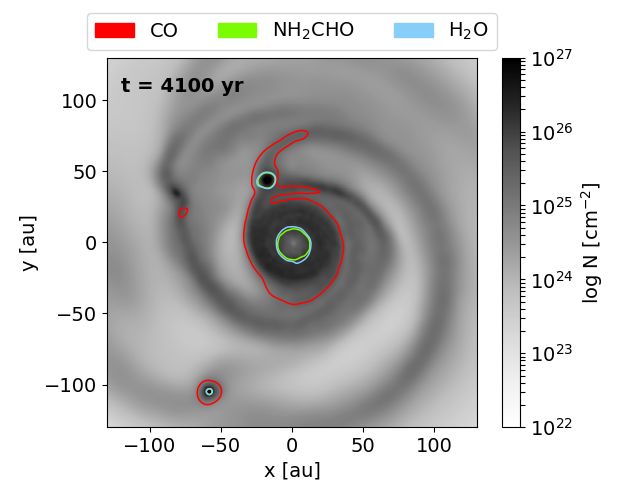}
\includegraphics[height=0.3\textwidth, trim={12.3cm 0cm 0.5cm 1.05cm}, clip]{snowline_063.png}
\caption{Evolution of the total column density of the disc (grey-scale) overlaid with the snow lines of CO (red), NH$_2$CHO (green) and H$_2$O (blue). The snow lines quickly deviate from the expected concentric ring structure due to the dynamic evolution of the disc.\vspace{0.5 cm}}
\label{snowline}
\end{figure*}

Fig. \ref{snowline} presents the evolution of the column density of the disc overlaid with contours showing the location of the mid-plane snow lines for CO, H$_2$O and NH$_2$CHO. 
The snow lines are defined by the radii at which the gas-phase column density of a species is equal to that of the column density of the corresponding ice-phase species, i.e. $N_{\rm gas} = N_{\rm ice}$ \citep[see Sect.~4.2 of][]{ilee2017}.
The CO snow line is much larger than the one for H$_2$O because of the larger desorption temperature for the latter. Since formamide is desorbed under similar conditions to water (their binding energies are similar, E$_{\rm D}$= 5600\,K for H$_2$O vs. E$_{\rm D}$= 6300\,K for NH$_2$CHO, see Table \ref{tableDes}), their snow lines are practically the same. We note that the formamide/water snow line is only present for the hottest fragments (George and John; middle panel of Fig. \ref{snowline}), because the amount of gas-phase formamide (or water) in Paul and Ringo is not high enough to match the amount of this molecule on the surfaces of dust grains.

\subsection{Radial distribution of formamide in the fragments}\label{sec:fragments}

\begin{figure*}[!ht]
	\centering
	\includegraphics[width=0.49\hsize]{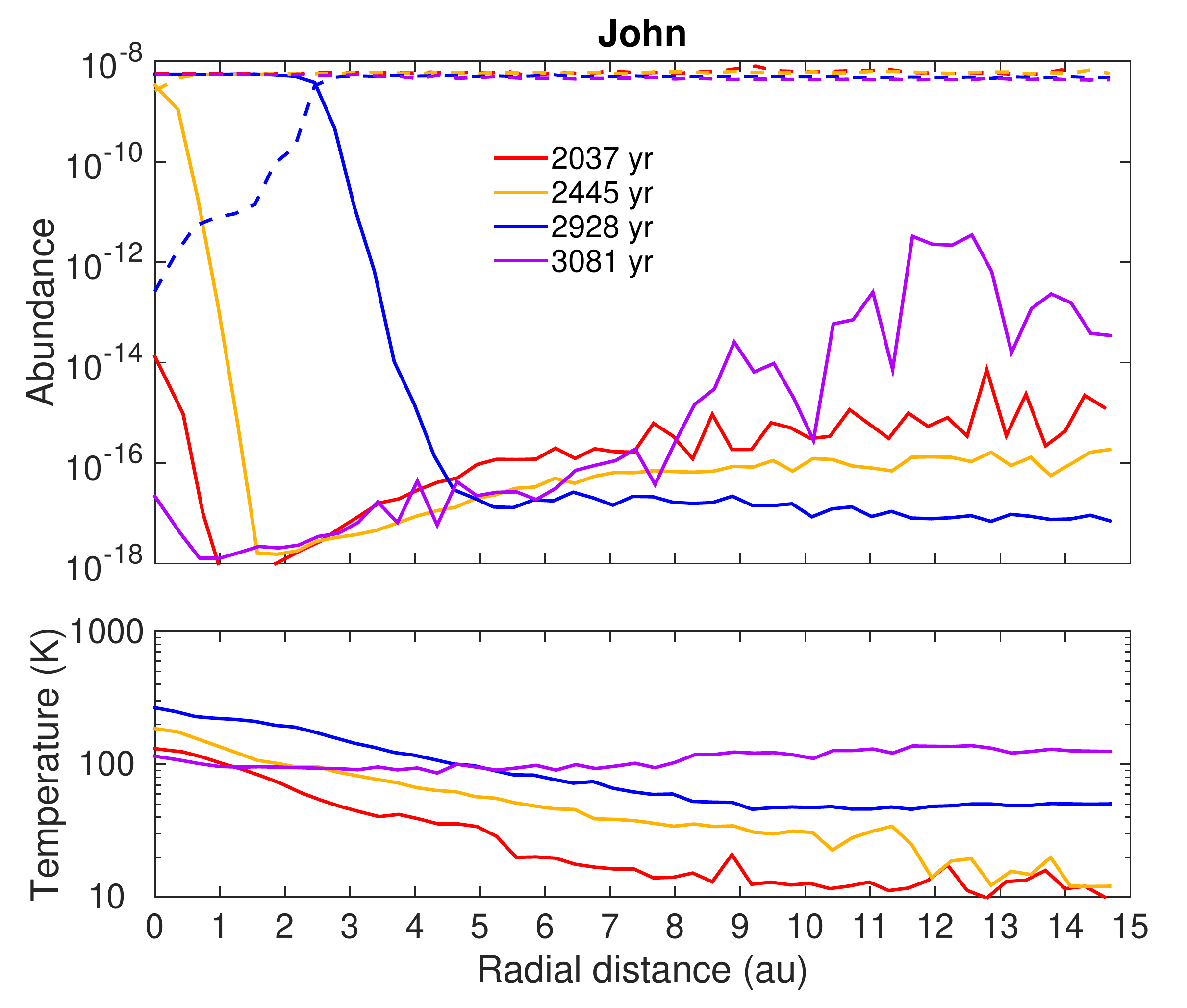}
	\includegraphics[width=0.49\hsize]{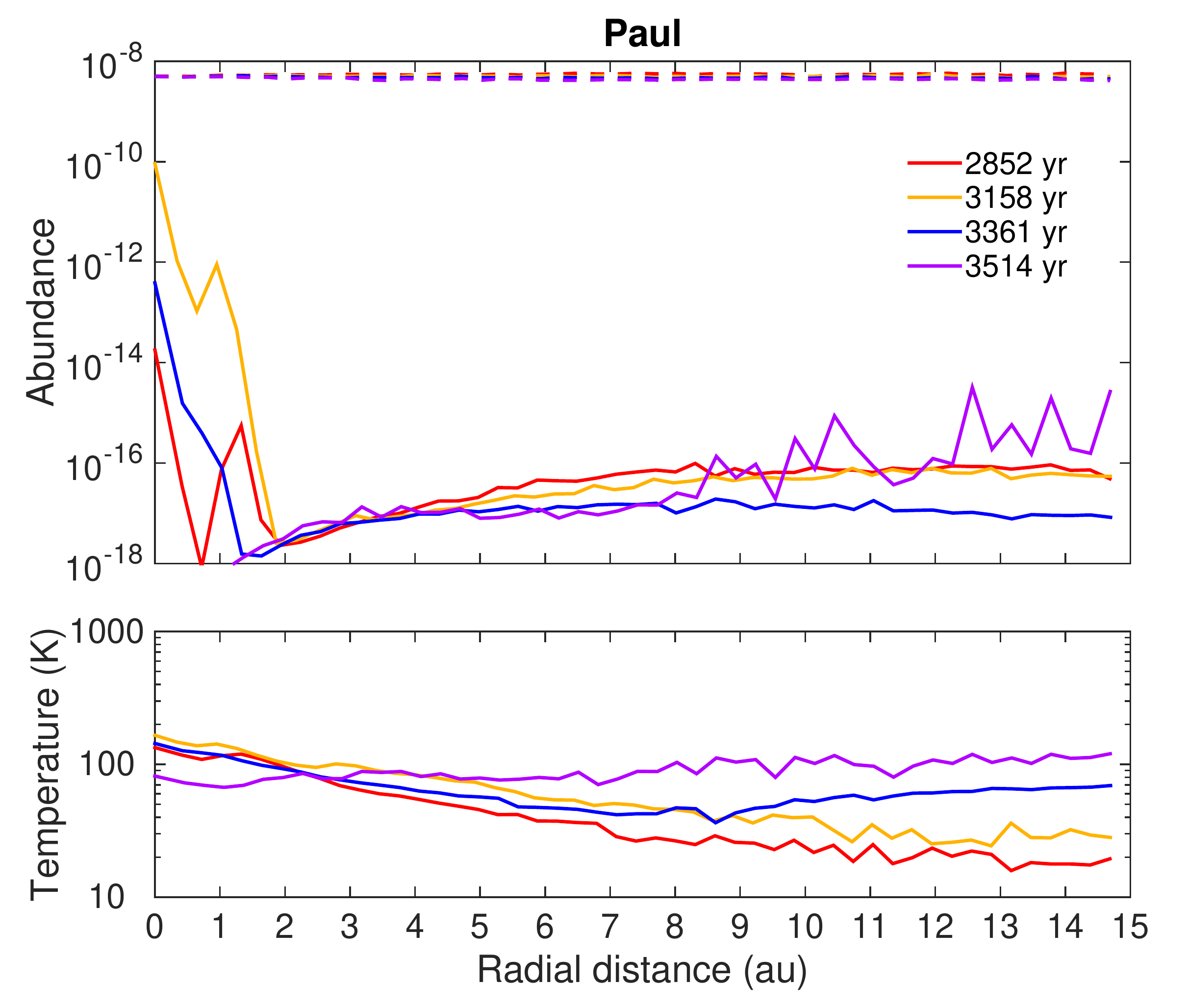}\\
	\includegraphics[width=0.49\hsize]{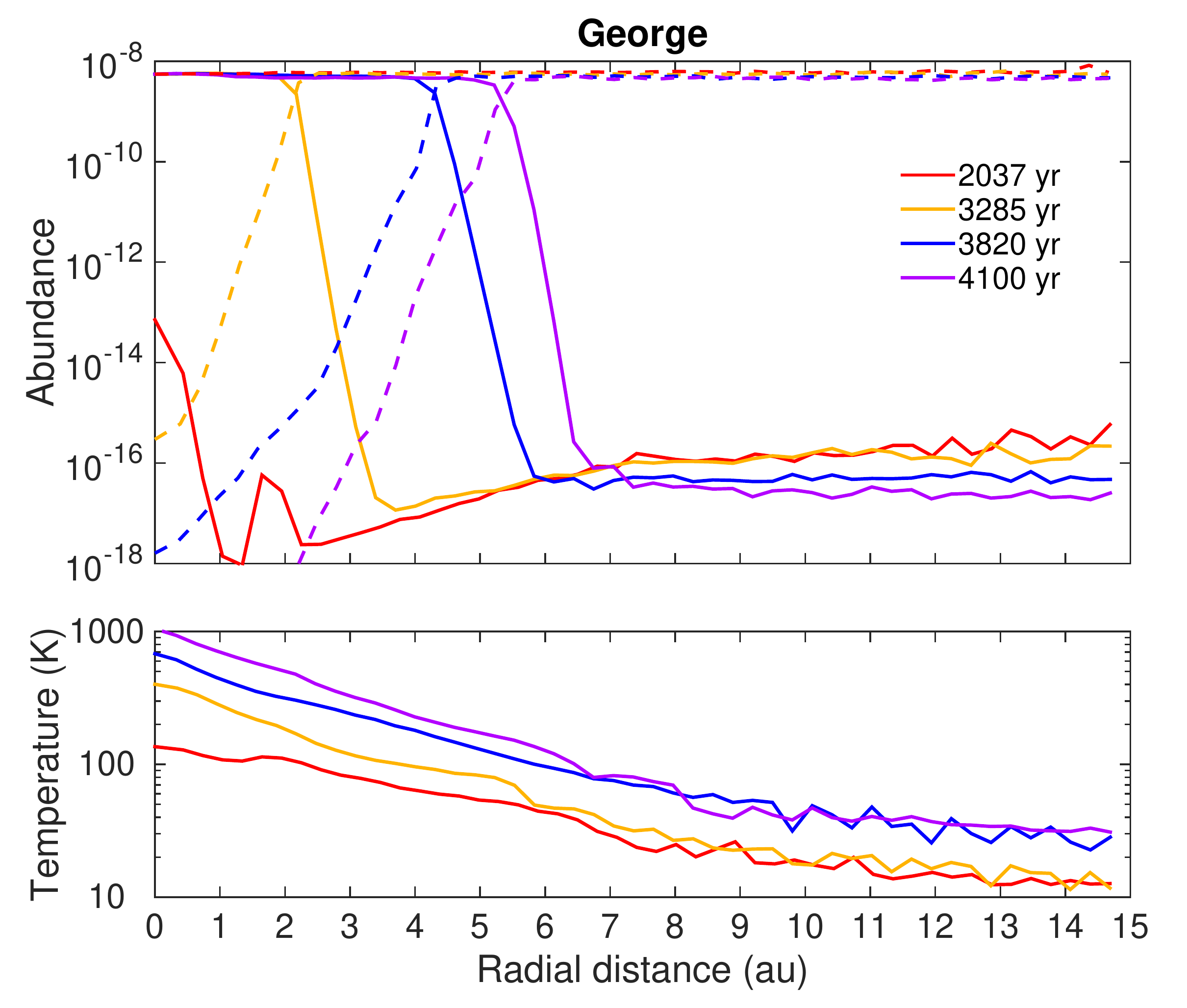}
	\includegraphics[width=0.49\hsize]{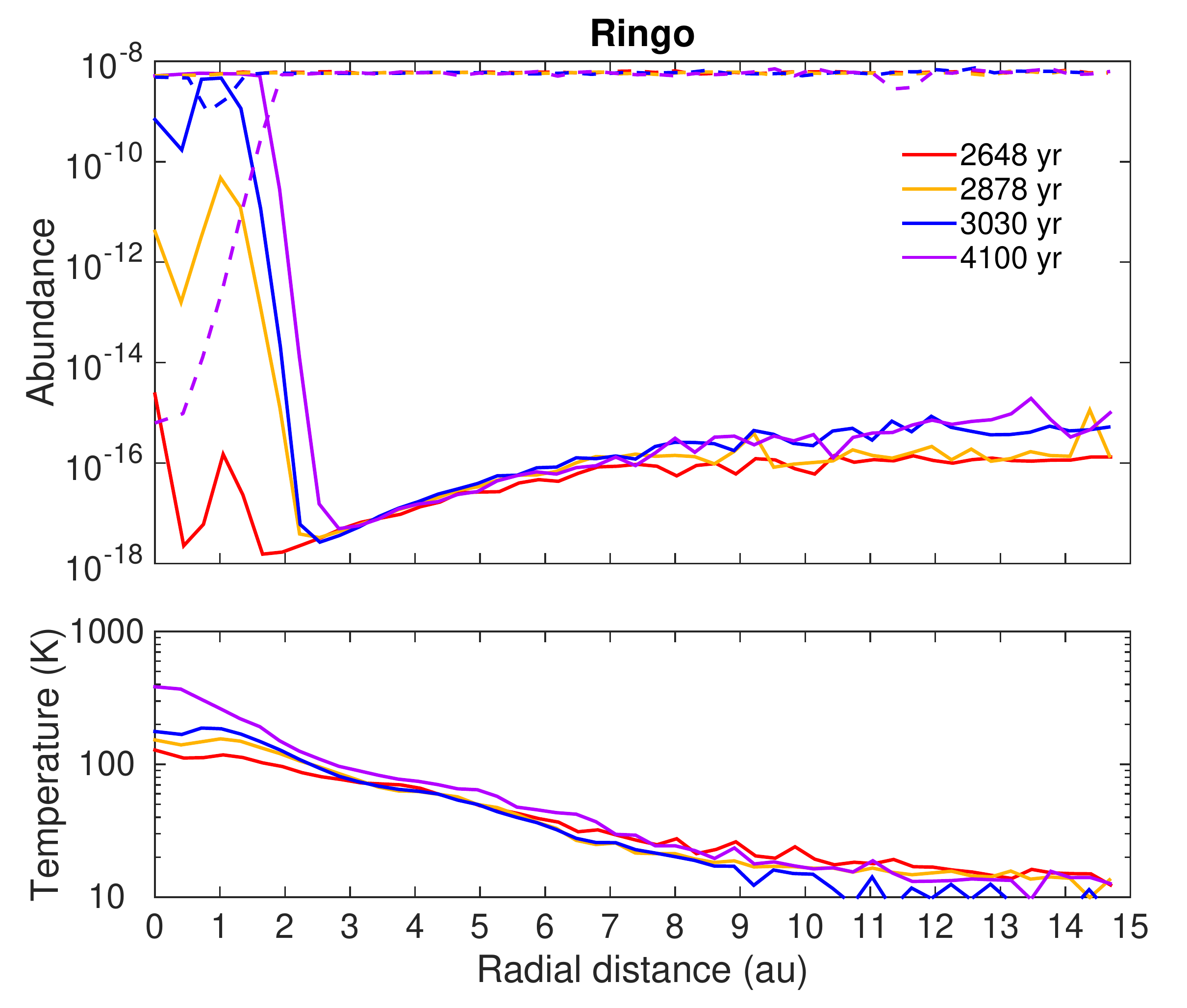}
	\caption{Median abundance of formamide in the gas-phase (full lines) and on the grain surface (dashed lines) as a function of the radius of each fragment: John (top left), Paul (top right), George (bottom left) and Ringo (bottom right). The lower panels show the median temperature profile around each fragment. Each colour represents a different time, comprised between the formation and destruction time (if any) of each fragment. \vspace{0.5 cm}}
	\label{allinone}
\end{figure*}

Fig. \ref{allinone} presents the median abundance of formamide for the four fragments as a function of distance from the fragment's centre for different times.  These  abundances are calculated taking into account the median of all abundance values within shells of 0.2\,au in width, with 50 shells covering a radial distance of 10\,au around each fragment). The time range for each fragment is taken between their formation time and the end of the simulation (or destruction time where appropriate, i.e. 2037--3081\,yr for John; 1630--3514\,yr for Paul; 2037--4100\,yr for George and 2445--4100\,yr for Ringo -- see also Table 2 of \citealp{ilee2017}).

\smallskip

For George, the radius of gas-phase formamide increases with time, to be $\sim$7\,au at the end of the simulation. This is due to the increasing temperature of the fragment (up to temperatures of 1100\,K), caused by its inward migration and increasing size with time. For Ringo, the radius of gas-phase formamide stays rather constant at around $\sim$2.5\,au. Indeed, Ringo has a relatively constant orbital radius as a function time, thus its radial temperature profile remains the same with time (T$\geq$100\,K for R$\lesssim$2.5\,au).  The temperature in Paul is much lower than the other fragments (150\,K vs. 300-1100\,K, at most), hence the majority of formamide remains frozen onto dust grains. Indeed the gas-phase abundance in Paul is less than a few 10$^{-10}$ within 0.5 au, corresponding to $<$1\% of the total formamide abundance of 10$^{-8}$ available on the grains.

\section{Discussion}

\subsection{Comparison with cometary abundances of formamide}\label{sec:comets}

In Section \ref{results}, we found that formamide is fully frozen on the surface of dust grains soon after the start of the simulation in the outermost regions of the disc and outside the fragments, as a result of their low temperatures and high densities (T $\leq$ 10 K and n(H$_2$) $\geq$ 10$^{10}$$\,$cm$^{-3}$). The solid-phase abundance of formamide in these regions, therefore, does not change significantly during the remaining evolution of the disc.  In a similar way, comets are thought to be formed from pristine material at the very edge of protoplanetary discs which do not undergo significant chemical processing. Therefore, the solid-phase abundance of formamide in our model at timescales $\sim$850 yr from the start of the disc simulation should be representative of such a situation.  In order to test this, we gathered observationally determined abundances of formamide and related species for three long period comets -- C/1995 `Hale-Bopp', C/2012 F6 `Lemmon' and C/2013 R1 `Lovejoy' (see \citealt[][and references therein]{biver2014, bockelee-morvan2000, le_roy2015}).  Table \ref{compare_abund} shows these abundances (relative to H$_2$O) alongside abundances measured in the outer disc regions early in the disc simulation.  We find that these abundances agree to within a factor 5 for all species, and note an excellent agreement between the modelled and observed values of formamide.  As such, our results show that even massive, unstable discs which undergo gravitational fragmentation can possess cold outer regions within which the abundance of complex organics can be similar to those observed in long period comets in the Solar System.

\begin{table}[!ht]
	\centering
	\caption{Solid-phase (ice) abundances in the outer parts of the disc model and compared to observed abundances in long-period comets. All values are relative to water (in percent).
	\label{compare_abund}}
	\begin{tabular}{lcc}
		\hline\hline
		Species	&	Model (\%)	&	Observed\tablenotemark{1} (\%)\\
		\hline
		NH$_2$CHO		&	0.02		& 0.01--0.021\\
		NH$_3$			&	1.08		&	0.61--0.7\\	
		H$_2$CO			&	0.23		&	0.7--1.1\\
		CO				&	9.65		&	4.0--23.0\\	
		CO$_2$			&	27.4		&	6\\
		CH$_4$			&	1.65		&	0.6--0.67\\	
		CH$_3$OH		&	1.00		&	1.6--2.6\\
		HCN				&	0.766	&	0.14--0.25\\
		\hline
	\end{tabular}
	\tablenotetext{1}{Long-period comets: C/1995 O1 (Hale-Bopp), C/2012 F6 (Lemmon) and C/2013 R1 (Lovejoy). Values taken from \citet{biver2014}, \citet{bockelee-morvan2000} and \citet{le_roy2015} (and references therein).}
\end{table}

\subsection{Formamide preservation around `cold' fragments}\label{paul_dust_settling}

In Section \ref{sec:fragments}, we show that most of the formamide content in Paul ($\sim$99\%) remains on the grain surface during its existence. The `cold' environment around Paul is a direct result of a smaller size and mass ($\sim$2\,au in radius and mass of 3.7\,$M_{\rm Jup}$) compared to the other three fragments (with radii between 2.5--7\,au and masses between 5.4--10.3\,$M_{\rm Jup}$). Paul does not survive long enough to further contract and heat up its surroundings (T$\leq$ 150\,K), so that water and many other molecules remain on icy grain surfaces. As proposed by \citet{ilee2017}, this effect may have important consequences for the formation of protoplanetary objects if the timescale for dust settling within the fragments is small enough. Indeed, \citet{ilee2017} inferred that the timescale for centimetre-sized silicate grains to settle to the centre of the fragment would be $1000-2000$\,yr, which is within the $\sim$2000\,yr of existence of Paul. Such dust settling might even occur more quickly, either due to icy grains possessing a high sticking co-efficient than bare grains \citep[e.g.][]{ehrenfreund_2003, wang_2005}, or via hydrodynamic instabilities within the fragments themselves \citep{nayakshin2018}.  Both effects would allow a higher proportion of solid material to sediment toward the centre of the fragment. If rocky protoplanetary cores are formed within these regions, then formamide (and other grain species, including larger complex organics) would be incorporated into them. Such rocky cores may go on to become terrestrial-like planets through e.g. tidal downsizing \citep{nayakshin2017, forgan2018}, sequestering primordial complex organics within them.

\subsection{Formamide preservation around ``hot'' fragments}\label{galilean_moons}

\begin{figure}[!tp]
	\centering
	\includegraphics[width=1.0\hsize]{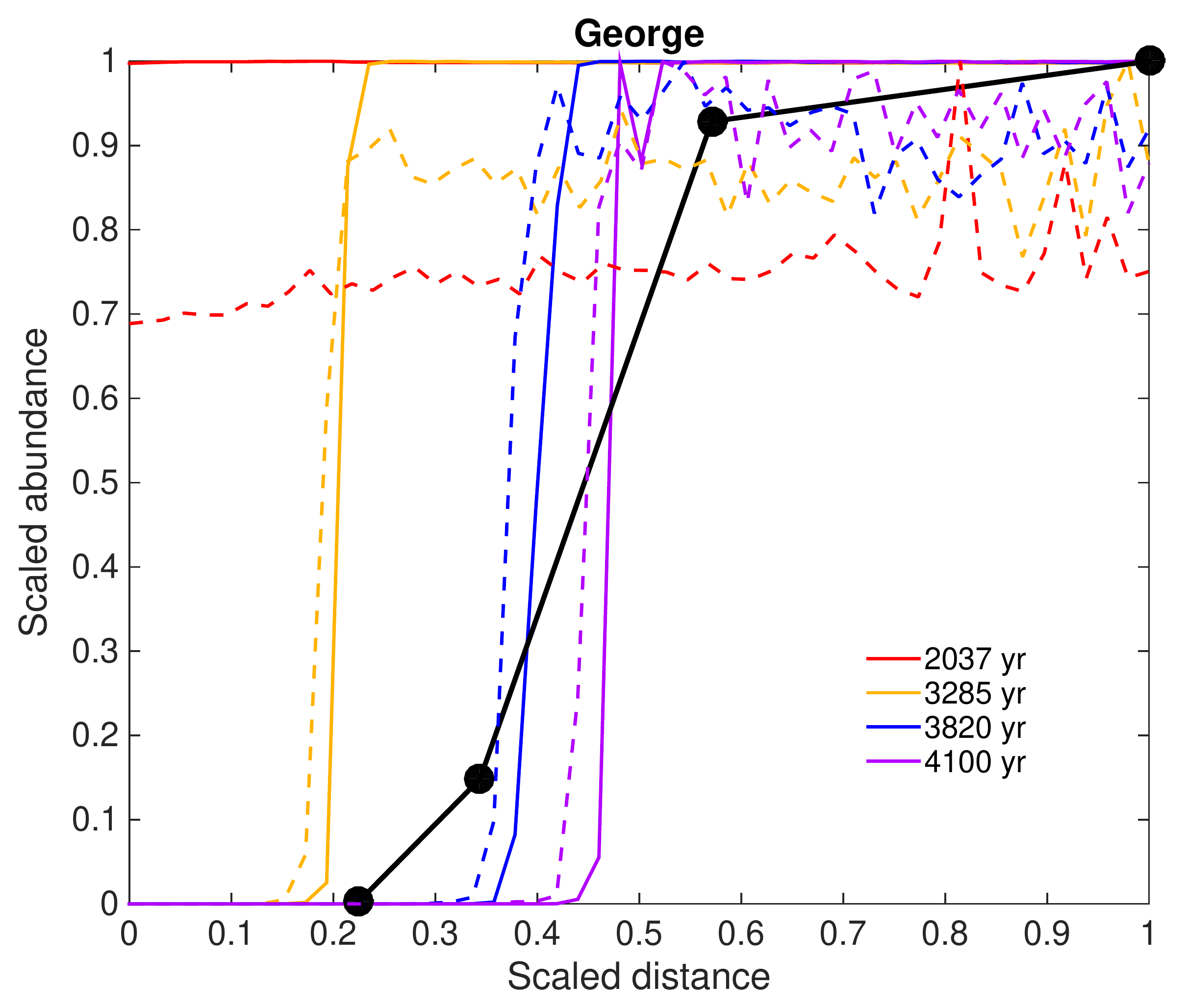}
	\caption{Scaled water (full coloured lines) and formamide (dashed coloured lines) abundances on the grain surface as a function of the radius of George. The scaled distance used for George is equal to 12\,au. Each colour represents a different time. The black line show the water content inside the four Galilean moons, as a function of their distance from Jupiter (scaled to the distance of Callisto).\vspace{0.5 cm}}
	\label{galilean}
\end{figure}

\smallskip

As shown in Section \ref{sec:fragments}, formamide is thermally desorbed into the gas-phase around hot fragments due to high temperatures. In the innermost regions of the fragments where T$\geq$500\,K, formamide is destroyed via gas-phase reactions for timescales longer than those considered in these simulations. While at larger radii, the temperature drops to levels at which formamide can be retained on dust grains (Fig. \ref{allinone}). Thus, while forming protoplanetary fragments would be devoid of solid-phase species, the circumfragmentary material around them would not be. Here we discuss the content of this circumfragmentary material, and the potential make-up of any planetary satellites that may go on to form from it.     

\smallskip

In Fig. \ref{galilean}, we present the distribution of water (in solid lines) and formamide (see dashed lines) in solid phase around George, the hottest surviving fragment in our simulations (different colours correspond to different timescales). We also overplot the solid water content measured in the Galilean moons: 46$-$48\,wt\% [percentage by mass] for Ganymede, 49$-$55\,wt\% for Callisto, 0\% for Io, and 6$-$9\,wt\% for Europa \citep{kuskov2001,kuskov2005}. The $x-$axis has been scaled to the radius of Callisto and to the radius for which the temperature around George equals the temperature of the disc (i.e. at a radius of 12\,au). From Fig. \ref{galilean}, we find that the fraction of solid water and formamide in our simulations increases with radius in a similar fashion to that observed for the solid water content in the Galilean moons. Indeed, the low temperatures ($\sim$40 K) and high water abundances found at the outer radius around George are consistent with an icy moon like Callisto being formed at low-temperatures, which could have prevented any melting of the ice \citep[while Io, Europa and Ganymede are completely differentiated bodies, Callisto presents a mantle without any separation between the ice and the rock material;][]{kuskov2005}. Therefore, formamide could also be retained in satellites formed around protoplanetary fragments from gravitationally unstable discs.  

\smallskip

We stress that this comparison is valid regardless of the formation scenario proposed for a Jupiter-like system \citep[e.g. whether the planet is formed via core accretion or gravitational instability;][]{canup2002,mosqueira2003}, because the expected shape of the radial temperature profile around nascent protoplanets formed via either mechanism are the same \citep[see][]{cleeves2015,szulagyi2017}. As such, water and formamide will always follow a similar radial behaviour as shown in Fig. \ref{galilean}.  Thus, while significant chemical processing can still occur in the later, Class II stages of protoplanetary disc evolution, our results demonstrate that such processing is not required to match the radial abundance profiles of Solar System satellites, and similarities can be in place at early times for objects formed via gravitational fragmentation of the disc.

\section{Summary}\label{ccl}

We have studied the formation and evolution of the key prebiotic molecule formamide in the context of a young, massive circumstellar disc which gravitationally fragments into bound protoplanets.  We find that formamide is located in the gas-phase in the innermost ($r<15$\,au) regions of the disc, and in three of the four fragments. It remains on the grain surfaces in the cold, outer regions of the disc where its abundance is consistent to that observed in long-period comets.  In one fragment, formamide remains on dust grains long enough to allow sedimentation, implying that rocky protoplanetary cores formed via, e.g., tidal downsizing, can inherit primordial pre-biotic material directly from the protostellar nebula.  In the circumfragmentary material, the radial abundance of solid formamide and water is consistent with the behaviour of solid water observed across Galilean satellites.  

\smallskip

In combination, our results show that when calculated explicitly, the chemical composition of protoplanets and protosatellites formed in gravitationally unstable discs can be similar to those expected from core accretion, and can even be consistent with the composition of objects within our own Solar System.

\acknowledgments

The authors are grateful to Serena Viti, Richard Nelson and Craig Agnor for valuable discussions, and to the anonymous referee for helpful suggestions to improve the manuscript. DQ and IJ-S acknowledge the support received from the STFC through an Ernest Rutherford Grant and Fellowship (grant numbers ST/M004139 and ST/L004801). JDI acknowledges support from the DISCSIM project, grant agreement 341137 under ERC-2013-ADG, and support from the STFC (grant number ST/R000549/1).  DHF acknowledges support from the ECOGAL project, grant agreement 291227, under ERC-2011-ADG. CH acknowledges that this project has received funding from the European Research Council (ERC) under the European Union's Horizon 2020 research and innovation programme (grant agreement No 681601). CH is a Winton Fellow and this research has been supported by Winton Philanthropies.

\software{\texttt{sphNG} \citep{bate_1995}; \texttt{UCLCHEM} \citep{holdship2017}}


\end{document}